\documentclass[3p,times,twocolumn]{elsarticle}

\usepackage{amssymb}

\usepackage[figuresright]{rotating}

\def\mathnew{\mathsurround=0pt}
\def\simov#1#2{\lower .5pt\vbox{\baselineskip0pt \lineskip-.5pt
       \ialign{$\mathnew#1\hfil##\hfil$\crcr#2\crcr\sim\crcr}}}
\def\simg{\mathrel{\mathpalette\simov >}}
\def\siml{\mathrel{\mathpalette\simov <}}

\def\Mesz{M\'esz\'aros\,}

\def\msun{M_\odot}

\def\vareps{\varepsilon}

\def\noind{\noindent}

\def\beq{\begin{equation}}
\def\enq{\end{equation}}
\def\bea{\begin{eqnarray}}
\def\ena{\end{eqnarray}}
\def\bec{\begin{center}}
\def\enc{\end{center}}

\def\blist{\begin{list}{$\bullet$}{\itemsep 0.0in \parsep 0.0in}}
\def\elist{\end{list}}
\def\bitem{\begin{list}{\arabic{enumi}.}{\usecounter{enumi} \itemsep 0.0in \parsep 0.0in}}
\def\eitem{\end{list}}
\def\cm{\hbox{~cm}}
\def\s{\hbox{~s}}
\def\erg{\hbox{~erg}}

\def\PeV{\hbox{PeV}}
\def\TeV{\hbox{~TeV}}

\def\MeV{\hbox{~MeV}}

\def\eV{\hbox{~eV}}

\def\part{\partial}

\def\Mpc{{\rm Mpc}}

\def\yr{{\rm yr}}

\def\m-pl{m_{Pl}}

\def\h75{h_{75}}
\def\Omh75{\Omega h^2_{75}}
\def\Omh70{\Omega h^2_{70}}

\def\G{{\rm G}}

\def\fun#1#2{\lower3.6pt\vbox{\baselineskip0pt\lineskip.9pt
  \ialign{$\mathsurround=0pt#1\hfil##\hfil$\crcr#2\crcr\sim\crcr}}}

\def\jcap{Jour. Cosmology and Astro-Particle Phys.\,}
\def\mnras{M.N.R.A.S.\,}

\def\apj{Astrophys.J.\,}
\def\apjl{Astrophys.J.Lett.\,}

\def\nat{Nature\,}

\def\prd{Phys.Rev.D\,}

\begin{document}

\begin{frontmatter}


\title{Multi-Messenger Signatures of\\
PeV-ZeV Cosmic Ray Sources}


\author[label1,label2,label3]{Peter \Mesz}
\author[label1,label3,label2]{Kohta Murase}
\author[label4]{Katsuaki Asano}
\author[label1,label3,label2]{Nicholas Senno}
\author[label5,6]{Di Xiao}
\address[label1]{Center for Particle and Gravitational Astrophysics, Pennsylvania State University, University Park, PA 16802, U.S.A.}
\address[label2]{Dept. of Astronomy \& Astrophysics, 525 Davey Laboratory, Pennsylvania State University, University Park, PA 16802, U.S.A.}
\address[label3]{Dept. of Physics, 102 Davey Laboratory, Pennsylvania State University, University Park, PA 16802, U.S.A.}
\address[label4]{Institute for Cosmic Ray Research, The University of Tokyo, 5-1-5 Kashiwanoha, Kashiwa, Chiba 277-8582, Japan}
\address[label5]{School of Astronomy and Space Science, Nanjing University, Nanjing 210093, China}
\address[label6]{Key Laboratory of Modern Astronomy and Astrophysics (Nanjing University), Ministry of Education, China}


\begin{abstract}

We discuss likely  sources of cosmic rays in the $10^{15}-10^{20}\eV$ range  and their
possible very high energy neutrino and gamma-ray signatures which could serve to identify
these sources and constrain their physics. Among these sources we discuss in particular
low luminosity gamma-ray bursts, including choked and shock-breakout objects, tidal
disruption events and white dwarf mergers. Among efforts aimed at simultaneous secondary 
multi-messenger detections we discuss the AMON program.

\end{abstract}


\end{frontmatter}


\section{Introduction}
\label{sec:intro}

Cosmic rays (CRs) up to ZeV $\equiv 10^{21}\eV$ are being detected by the Pierre Auger Observatory
and the Telescope Array. Being mainly charged particles, they are very hard to trace back to their 
original sources, since they are scattered by intergalactic and interstellar magnetic fields and
loose their original directional information. Even for ultra-high energy cosmic rays (UHECRs)
in the Greisen-Zatsepin-Kuz'min (GZK) range $\simg  6\times 10^{19}\eV$, the error circles are
of the order of a degree for protons, and more for heavier elements. On the other hand,
CRs are relativistic hadrons, and colliding with low energy or thermal protons and photons 
in their sources of origin, or along their path to the observer, they produce copious numbers of
secondary particles, which end up as neutrinos, $\gamma$-rays and $e^\pm$. The $e^\pm$ end
up producing $\gamma$-rays or lower energy photons of degraded directionality, while the secondary
$\gamma$-rays whose initial energy exceeds the $\gamma + \gamma_{EBL} \to e^+ +e^-$ pair-formation
threshold against the diffuse external background light (EBL) also loose most of their directionality,
producing a lower energy isotropic gamma-ray background (IGB) with a universal spectral shape
\cite{Berezinsky+75nucasc}, mostly in the $\siml \TeV$ gamma-ray range.

Any direct CR secondary $\gamma$-rays of energy $E_\gamma\siml 0.5-1 \TeV$, however,  depending
on the redshift of the source, can travel directly to the observer. Also CR-secondary neutrinos 
of any energy (and neutrinos in general) can travel directly to the observer, with negligible
interaction along the way, even from the highest redshift sources. Such direct CR secondary
$\gamma$-rays (or lower energy photons) and neutrinos are therefore a prime tool, and perhaps
the main if not only tool, to infer the source locations and allow follow-up observations with
other instruments, such as X-ray, optical or radio telescopes. Another possible channel is
gravitational waves (GWs), recently discovered form stellar mass binary black hole mergers
\cite{LIGO+16-gw150914disc,LIGO+16gw151226}, but the GW localization error boxes are extremely 
large for the foreseeable future, and so far it is unclear whether CRs are expected from such mergers. 
On the other hand, binary neutron star mergers are also expected to emit GWs which should be 
discovered by LIGO/VIRGO anytime soon, and these are thought to be related to short gamma-ray bursts 
(SGRBs), which could accelerate cosmic rays, e.g. \cite{Meszaros13cta}.  
Thus, there is a rich trove of different messenger (multi-messenger) particles which can provide 
information about not only the sources of cosmic rays, but also about the physics of the sources, 
providing clues or constraints about the acceleration process, the source and its direct environment, 
and the intervening medium between the source host and the observer.

\section{GZK UHECRs and below: Classical GRBs?}
\label{sec:grb} 

The UHECR spectrum has been measured by Auger \cite{Auger+15-augerspec} and TA in the range
$10^{17.5}\eV-10^{20.5}\eV$. Above $\sim 10^{18.5}\eV$ the CRs cannot be contained in typical 
galaxies such as ours, which means that these are guaranteed to be extragalactic, and also that
it is likely that the spectrum at these energies reflects the spectrum as they escaped form the 
accelerator. The spectrum in this range is roughly compatible with a slope close to $N(E)\propto 
E^{-2}$, aside from the feature called the ankle, one possible explanation for which might be a
Bethe-Heitler absorption, e.g. \cite{Aloisio+12galegal}. Below $\sim 10^{18}\eV$ the observed
spectrum is $\propto E^{-3}$ down to PeV, and below that it is $\propto E^{-2.7}$, significantly
steeper. However, below the ankle the CRs can be trapped in galaxies for long times, and judging 
from our own galaxy's energy dependence of  the diffusion coefficient, one would expect the
observed spectral slope to be significantly steeper than that of the produced spectrum, roughly
in accordance with observations if the produced spectrum were of slope roughly -2 at all energies,
e.g. \cite{Wick+04grbcr}. In fact, the energy production rate per unit volume in the universe
per decade of energy, $E^2 (dN/dE)$ at all energies between $10^9\eV$ and $\sim 10^{20}\eV$ 
has been estimated by \cite{Katz+13uhecr} to be approximately constant over these 11 decades 
of energy, $\sim 10^{44}\erg~\Mpc^{-3}\yr^{-1}$, which is comparable to the CR flux resulting in
the Waxman-Bahcall neutrino upper bound \cite{Bahcall+01bound}. 

IceCube has found a diffuse extragalactic neutrino background flux in the TeV-PeV range
at the WB level \cite{IC3+13pevnu2,IC3+15tevnu}, which however cannot be explained by ``classical" 
GRB internal shocks or other models \cite{Abbasi+12-IC3grbnu-nat,IC3+15grbnu4yr}, both due to 
time and location window non-agreements with electromagnetic detections, and due to over-predicted 
fluxes.  It is worth noting however that the original (also approximate) original calculation of 
VHE neutrinos from classical GRB internal shocks \cite{Waxman+97grbnu} predicted flux levels a 
factor $\sim 10$ below that later measured by IceCube, i.e. not in conflict, although later 
approximate estimates by other groups from similar models obtained higher values.  More exact
neutrino calculations \cite{He+12nugrb,Li12grbnu,Hummer+12nu-ic3} resulted in classical internal
shock \cite{He+12nugrb,Li12grbnu,Hummer+12nu-ic3} and photospheric \cite{Murase08grbphotnu,
Wang+09grbphotnu,Gao+12photnu,Murase+13subphotnu}
GRB flux levels significantly below the IceCube measured flux levels. This, however, does not
preclude the possibility that classical GRBs could still be sources UHECRs. This could plausibly
be the case in the above models if the pion production (i.e. $p\gamma$) efficiency were low,
e.g. if the shocks or dissipation regions accelerating the CRs were moderately larger than 
usually assumed, e.g. \cite{Murase+12reac,Asano+14grbcr}.

Alternatively, it may be that classical GRBs provide a solution only for part of the PeV-ZeV spectrum
For example, one can show \cite{Asano+16grbcrturb} that even if classical GRBs do not explain 
the IceCube neutrinos they could be the sources of the GZK cosmic rays. In this particular
twist of the classical GRB model, it is assumed that the MeV gamma-rays of GRBs arise in the 
GRB photosphere (as recent models argue, e.g. \cite{Rees+05photdis,Beloborodov10pn}). Shocks 
must inevitably occur outside the photosphere, if nothing else then when the ejecta is decelerated
by the external medium. These  shocks can accelerate cosmic rays, and if this occurs via a
2nd order Fermi process (since there is turbulence behind the shocks, e.g. \cite{Duffell+13grbes}), 
this process produces a CR energy spectrum which is flatter than the conventional -2 slope of 
1st order Fermi.
\begin{figure}[ht]
\includegraphics[width=0.5\textwidth,height=2.0in,angle=0.0]{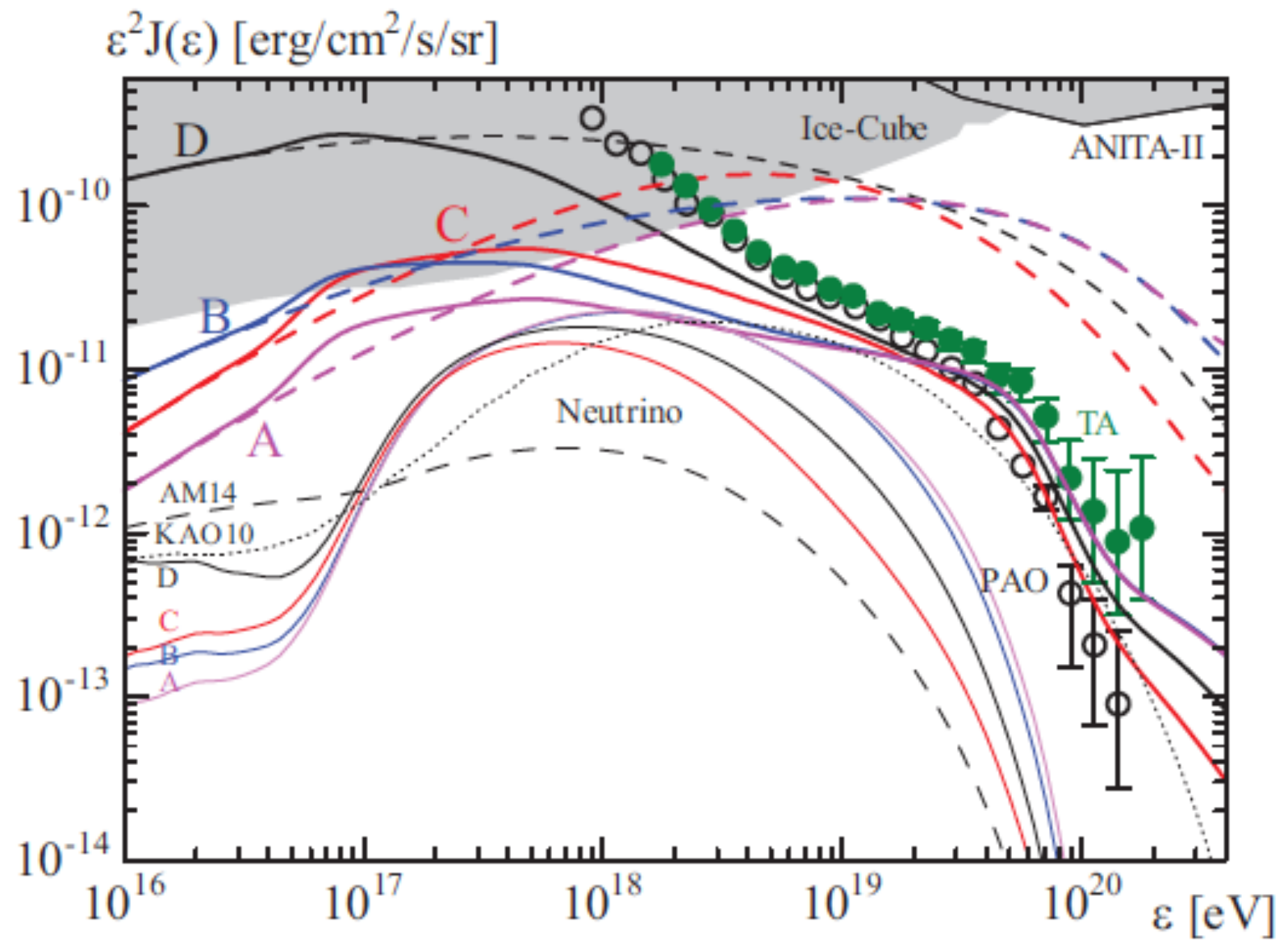}
\caption{Diffuse UHECR spectrum for several GRB 2nd order Fermi models \cite{Asano+16grbcrturb} 
are shown by thick solid lines.
The same neglecting photo-meson production and Bethe-Heitler pair formation are in thick dashed lines.
Auger data: open circles, TA data: green filled circles. Thin solid lines: all-flavor cosmogenic neutrino
flux from these models . The thin dashed and dotted lines are cosmogenic neutrinos from other calculations. 
Upper gray shaded area: the IceCube limits. 
}
\label{fig:Asano+16crfg4}
\end{figure}
With this, the Auger cosmic ray spectrum in the $10^{18}-10^{20}$ eV range can be explained without 
violating the IceCube limits, see Fig. \ref{fig:Asano+16crfg4}.  This is because the CR energy is 
mostly concentrated at the upper end of the spectrum (above the ankle, which is thus explained).
And, since the slope is flatter than -2, there is significantly less CR energy in the 10-100 PeV 
range, which thus produces a PeV neutrino flux below that observed by  IceCube. Also, since the 
MeV photons coming from much further below than where the CRs are accelerated, the $p\gamma$ 
efficiency is much lower, since the photon flux is much more diluted than in one-zone models. 
The flatter spectral slope also eases considerably the baryon load needs, which thus is
easily compatible with the total energetic budget of GRBs.

\section{PeV-EeV CRs from Not-so-Classical GRBs?}
\label{sec:choked}

There is a class of so-called low luminosity GRBs (LLGRBs) which, because of its lower photon
luminosity has much fewer well-studied examples, all of them so far at low redshifts. 
Their general characteristics appear similar to those of classical high luminosity GRBs, in the sense
that their MeV photons spectrum indicates that electrons are accelerated to non-thermal relativist 
energies, presumably by shocks or dissipation in a relativistic outflow, albeit probably of lower 
Lorentz factor. However, their occurrence rate per unit volume appears to be at least an order of
magnitude higher than for their classical counterparts \cite{Howell+13-llgrb}, and if they accelerate
CRs they could be a significant contributor to the TeV-PeV neutrino flux, e.g. \cite{Murase+06llgrbnu,
Gupta+07nullgrb,Liu+11-llgrbcr}. The CRs responsible for this would be at least in the $\siml 100 \PeV$ 
energy range.

One can think of three possible GRB life histories which could give rise to LLGRBs, depending 
on how much energy the relativistic jet received in its infancy, and for how long, in the basic 
collapsar scenario. That is, a massive star's core collapses, a black hole (or  perhaps
temporarily a magnetar) forms, infall matter accretes and is ejected in a relativistic jet.\\
\noind
{\bf (1)} If the jet is accretion-fed, but not generously enough, or for not long enough, it stalls before 
it can emerge from the envelope: it is a {\bf choked jet} \cite{Meszaros+01choked}. \\
{\bf (2)} If the jet is fed a little longer, or a little more momentum is pumped into it so that it can just 
reach the stellar envelope surface or a bit beyond, the shock ahead of the jet may break out of the
envelope and the surrounding wind (a {\bf shock-breakout}), producing a weak, soft GRB-like EM 
radiation, e.g. \cite{Campana+06-060218,Waxman+07-060218}.\\
{\bf (3)} If the jet is fed just enough and for long enough that it can emerge completely from the 
stellar envelope and wind, it appears as an {\bf emergent}, EM-manifest LLGRB, again with 
a weaker, softer gamma-ray spectrum\footnote{
The original case is of course the well- and long-enough fed classical high luminosity GRB discussed 
in the previous section, of which thousands have been observed, and which may seem pushy, flashy and 
overfed, compared to their more modest and underfed but more abundant LLGRB relatives}.\\
Of these three LLGRBs, the first is expected to be only detectable through VHE neutrinos produced by
internal shocks or dissipation in the stalled, sub-surface jet, e.g. \cite{Meszaros+01choked,
Horiuchi+08choked, Murase+13choked, Varela+15choked, Fraija15pevchoked}. The  second and
third sub-scenarios could have a neutrino precursor from sub-surface shocks before the jet
emerges \cite{Meszaros+01choked}, followed by a LLGRB  EM burst. 
The $\gamma$-ray and optical-UV light-curves and spectral properties of the shock breakout LLGRBs
have been discussed, e.g.  by \cite{Waxman+07-060218,Waxman+03uncorking,Li07,Katz+11nusn-break,
Barniol+15llgrbag,Nakar15-llgrb,Irwin+16breakout}, and the neutrino properties by, e.g. 
\cite{Katz+11nusn-break,Kashiyama+13breakout,Giacinti+15nusn}, general reviews of shock breakout 
theory being given in, e.g. \cite{Irwin+16breakout, Waxman+16shockbrk}. 

\begin{figure}[ht]
\includegraphics[width=0.5\textwidth,height=2.0in,angle=0.0]{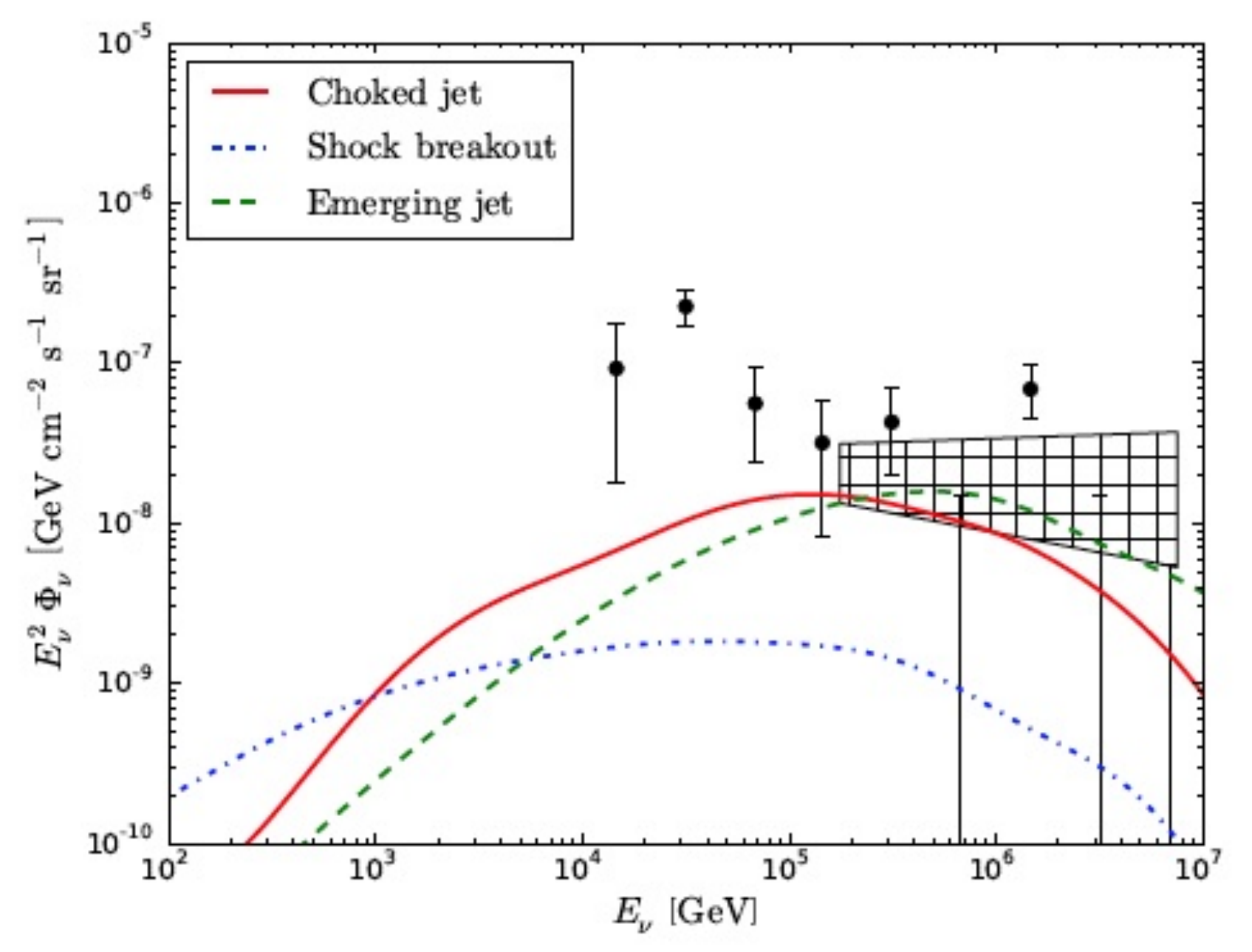}
\caption{
All-flavor diffuse neutrino fluxes expected from low-luminosity GRBs of three types: choked jets 
(orphan neutrinos, in red); precursor and shock-breakout neutrinos (blue); and prompt emergent 
jet LLGRB neutrinos (dashed). Overlaid are the IceCube data points based on the combined analysis 
\cite{Aartsen+15-IC3nubkggmaxlik} and up-going neutrino analysis (shaded, A. Ishihara \& IceCube 
collaboration, in talks at TeV Particle Astrophysics, 2015).  From \cite{Senno+16hidden}.  }
\label{fig:Senno+16hiddenfg3}
\end{figure}

A recent comparative study of the neutrino properties and the expected diffuse neutrino background 
from all three types of LLGRBs is in \cite{Senno+16hidden}, see Fig. \ref{fig:Senno+16hiddenfg3}.
This calculation \cite{Senno+16hidden} shows that a combination of choked jet, shock-breakout
and emergent LLGRBs is able, for conservative parameters, to  explain the observed IceCube 
diffuse neutrino flux. Furthermore, it does so  without violating either the Fermi observations 
nor the (classical GRB) stacked neutrino analyses, because the low luminosity of the breakout
or emergent LLGRBs (the majority of whom are at high redshifts) is too low to trigger Swift
or Fermi, while the choked jets are by definition gamma-silent. Thus, they are EM-hidden sources,
a property which seems necessary \cite{Murase+16hidden} for explaining the neutrino background 
while satisfying the Fermi isotropic gamma-ray background, 

\section{Tidal Disruption, UHECRs and Neutrinos}
\label{sec:tde}

Most if not all typical galaxies have a massive black hole (MBH) at the center, e.g. in our 
Milky Way the MBH mass is $\sim 2\times 10^6\msun$, and in the much rarer AGNs the masses can 
reach $\simg 10^9\msun$. The number density of stars in the central parsec around the MBH in
typical galactic nuclei is much larger than the average density in the disk, and occasionally
the stellar orbits can pass so close to the MBH that, if the star is not directly swallowed,
it is disrupted, a phenomenon known as a tidal disruption event (TDE).
For a star of mass $M_\ast$ and radius $R_\ast$ in the field of a BH of mass $M_{MBH}$ 
and Schwarzschild radius $R_{MBH,S}=2GM_{MBH}/c^2=3\times 10^{11}M_{MBH,6}\cm$ there is a tidal radius
$r_t\sim (2 M_\ast/M_{MBH})^{1/3}R_\ast \sim 5\times 10^{12}M_{MBH,6}^{-1/3}(R_\ast/R_\odot)
(M_\ast/\msun)^{1/3}\cm$. For MBHs more massive than about $7\times 10^7\msun$, solar type
stars are swallowed whole (since their tidal radius falls inside the MBH Schwarzschild radius),
but for lower mass MBHs stars smaller than solar type and white dwarfs are disrupted before being
swallowed. In these cases, about half the material disrupted near the periastron goes out on 
parabolic orbits, while the other half remains bound, falling back towards the MBH \cite{Rees88tid}.
In such TDEs, the stellar compression and shock during the initial passage around the periastron 
produces a prompt X-ray and gravitational wave signal, e.g.  \cite{Kobayashi+04tid}, while the 
subsequent fall-back material gives rise to further shocks leading to an optical and X-ray 
light curve of luminosity scaling with the mass accretion rate, $L\propto t^{-5/3}$. Many such 
TDEs are detected as transient X-ray and optical events, e.g. \cite{Komossa15tderev}. The 
fall-back is in fact complicated, although retaining an approximate $t^{-5/3}$ overall behavior.
There are multiple intersecting stream shocks followed by a slow circularization of the settling 
material \cite{Shiokawa+15tid,Piran+15tidcirc}. 
Even ignoring such complications, it was realized that in any case
the gas must undergo shocks, if nothing else as it joins an accumulating accretion disk, and 
such shocks may result in UHECR acceleration \cite{Farrar+09-crtid,Farrar+14tdeuhecr}. The rate, 
while poorly known, is approximately right for explaining the Auger UHECR flux if the CR 
acceleration efficiency is significant. 

In some TDE events one detects an initial GRB-like $\gamma$-ray transient, followed by a much longer 
X-ray decay which, at least in part, follows roughly the expected $t^{-5/3}$ law. The properties 
of the initial GRB-like behavior followed by the TDE-like behavior suggest that in a fraction 
of TDEs the accretion results initially in a relativistic jet \cite{Burrows+11-1644tid}.
Since the galactic bulge environments can be very gas-rich, and the disruption may be preceded 
by initial, gradually shrinking, matter-shedding periastron passages, an optically thick wind 
may be created before the jet is launched, which would thus be obscured by this pre-ejected wind  
\cite{Metzger+16tid}.

Relativistic jets launched under such an umbrella of a dense gaseous outflow resemble the
GRB jets propagating initially into the dense progenitor stellar envelope. This could result 
in a choked jet type of phenomenon, with internal and termination shocks in the jet leading to 
electron and proton acceleration, from which the $p\gamma$ neutrinos could escape 
\cite{Wang+16tdecrnu}, while the GRB-like leptonic radiation could be absorbed or thermalized 
by the dense wind. Thus. these sources would not be expected to trigger Swift or Fermi, i.e. 
for practical purposes they would be EM-hidden neutrino sources.

\begin{figure}[ht]
\includegraphics[width=0.5\textwidth,height=2.0in,angle=0.0]{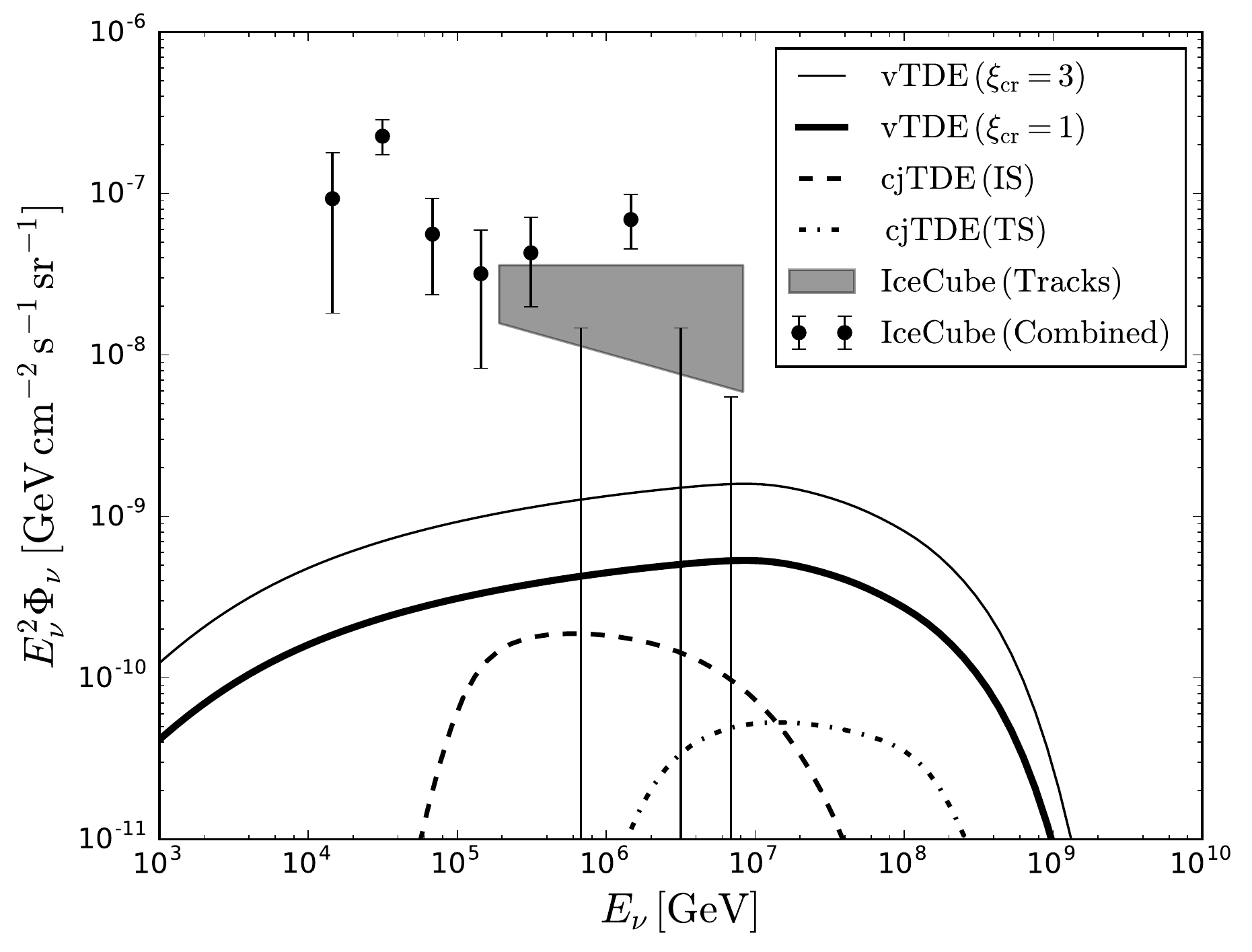}
\caption{All-flavor neutrino diffuse flux from shock breakout ($\nu$-TDE) and choked-jet (cjTDE) 
from $p\gamma$ interaction against jet head, IS synchrotron and envelope photon fields. 
The isotropic equivalent photon luminosity is $L_\gamma =10^{48} \erg \s^{-1}$ for v-TDE and $L_\gamma = 5 \times 10^{44} \erg \s^{-1}$ for cjTDE.
The overall $p\gamma$ neutrino diffuse flux is at the lower limit of the detection threshold, 
shown by the gray shaded region and data points with error bars. From \cite{Senno+17tde}.
}
\label{fig:Senno+17tdefg6}
\end{figure}

A more detailed consideration of the jet and wind conditions \cite{Senno+17tde} indicates 
that both choked jet and shock breakouts are expected, numerical calculations of the resulting 
diffuse neutrino spectra from such TDE jets in dense galactic bulge winds being shown in
Fig. \ref{fig:Senno+17tdefg6}, where choked jets and breakouts are plotted separately.
The total predicted diffuse flux is below the current IceCube detection threshold, although
there are substantial uncertainties in the rates. Even so, above 10 PeV such TDEs could start being
substantial contributors to the neutrino spectrum, the corresponding CRs being in the 
$\simg 0.2$ EeV.  Similar calculations appearing almost simultaneously are 
\cite{Dai+16tdeic3nu,Lunardini+17tde}.

\section{White Dwarf Mergers as CR/$\nu$ Sources}
\label{sec:wd}

White dwarf (WD) binaries are a common occurrence, and the tighter binaries will eventually merge
in less than a Hubble time, initially under the action of magnetic and tidal torques, and later due 
to gravitational wave emission. The merger rate has been estimated to be close to that of SN Ia, 
and in fact they have been proposed as a possible mechanism for SN Ia explosions. The conditions 
for such mergers to result in SNe Ia are debated, and not all mergers may lead to such explosions.
Even so, WD mergers  should result in optically thick magnetic outflows, which could lead to 
interesting bright optical transients \cite{Beloborodov14wdmerg}.
Numerical simulations  indicate that the merger is expected to result in a 
central core and a surrounding disk with a viscous accretion time $t_{visc}\sim 10^4\s$ and 
strong magnetic fields of order $B\sim 10^{10}-10^{11}\G$. 

\begin{figure}[ht]
\includegraphics[width=0.5\textwidth,height=2.0in,angle=0.0]{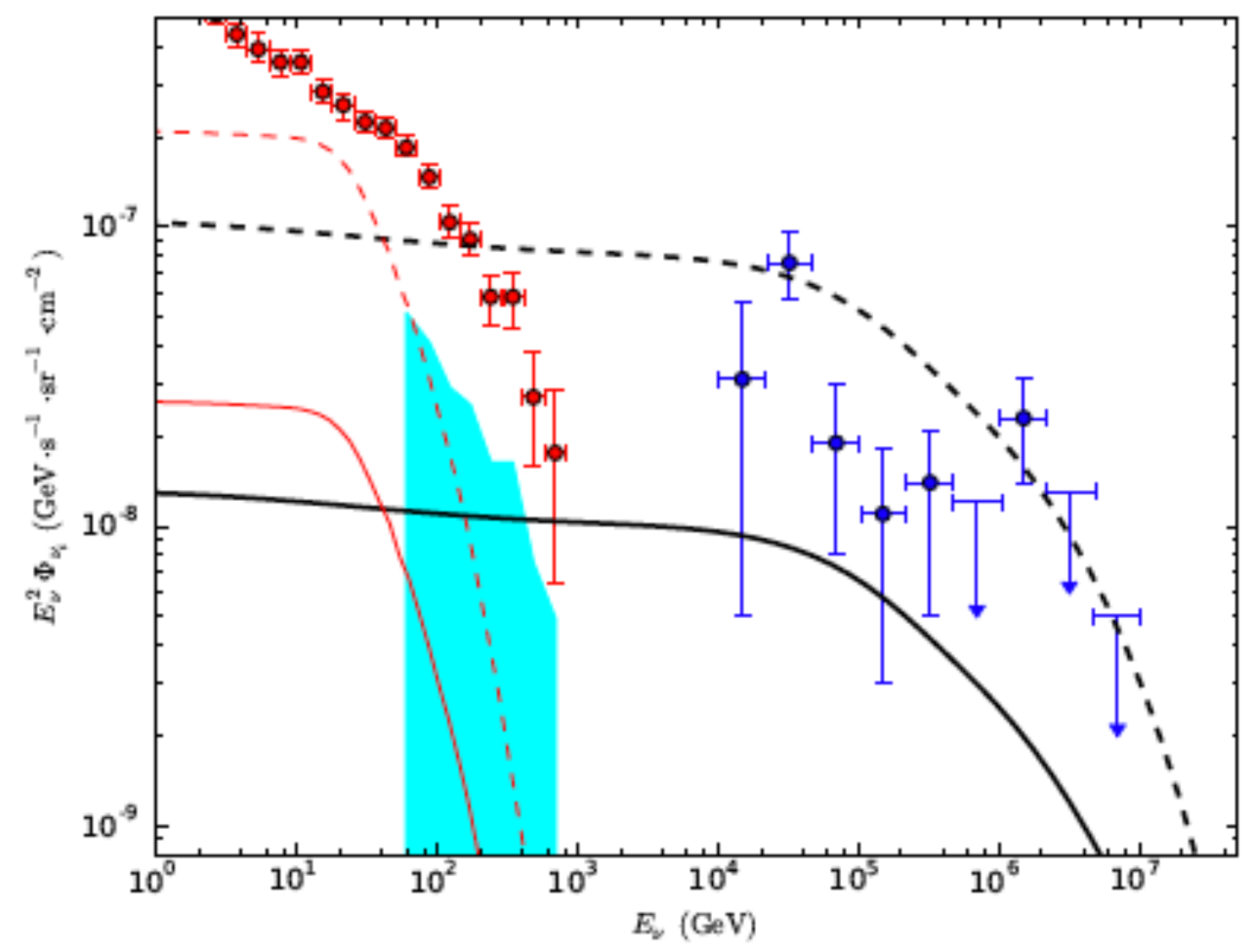}
\caption{
The ``optimistic" diffuse gamma-ray flux of the WD
merger scenario, which shows that the predicted gamma-ray flux
should be far below the extragalactic gamma-ray background measured
by Fermi-LAT (red data points) (Ackermann et al. 2015).
The cyan area shows the allowed region for the non-blazar gamma-ray
flux in Fermi Collaboration (2016). Only the absorption
effect is included in this figure, although the Bethe-Heitler
pair production in the ejecta is also likely to be important. The
thin red solid line is for the optimistic injection with $Q_{inj} =
10^{45} \erg\Mpc^{−3} \yr^{−1}$ and thin red dashed line is for the even more
optimistic case of 8 times higher. The thick black lines are the
neutrino fluxes, correspondingly.  From \cite{Xiao+16wdmnu}.  
}
\label{fig:Xiao+16wdfg5}
\end{figure}

The resulting  magnetically dominated outflows could have a luminosity $L_B\sim 10^{44}-10^{46}
\erg\s^{-1}$ with a total energy output of $\vareps_B\sim L_B t_{visc} \sim 10^{48}-10^{50}\erg$
\cite{Xiao+16wdmnu}. The flow initially is very optically thick, expanding at the escape velocity,
and magnetic reconnection is inhibited until a radius where the photon diffusion time becomes 
shorter than the dynamic time, where photons start to diffuse out, and magnetic reconnection 
can begin occurring. Reconnection can lead to particle acceleration on a timescale comparable to
that of Fermi processes, 
and in the magnetic fields beyond the diffusion radius this can accelerate protons to energies
$E_p\simg 100\PeV$. The $p\gamma$ interactions with the flow's thermal and synchrotron photons
as well as $pp$ interactions lead to VHE neutrinos in the $\siml$ few PeV range.
The merger rates have uncertainties, as well as the physics of the outflow and reconnection.
Bracketing these uncertainties  between an optimistic and a pessimistic case, the predicted 
diffuse neutrinos fluxes \cite{Xiao+16wdmnu} are shown in Fig. \ref{fig:Xiao+16wdfg5}.

While the Thomson scattering optical depth at the diffusion radius is still large, 
$\tau_T \sim c/V_{dyn}\gg 1$, the high energy $\gamma$-rays see a lower Compton cross section,
but they are also subjected to $\gamma\gamma$ annihilation and Bethe-Heitler matter absorption.
The resulting net effect gives and upper limit to the WD merger contribution to the diffuse  
isotropic gamma-ray background, also shown in Fig. \ref{fig:Xiao+16wdfg5} for the optimistic
and pessimistic cases. It is seen that these upper limits fall well inside  the level 
permitted by Fermi observations after subtracting contributions from  unresolved blazars.
Thus, these WD mergers can be considered another case of effectively EM-dark  sources, as far 
as not triggering satellite detectors looking for sudden increases of $\simg \MeV$ photons. 

Even for the pessimistic rate estimate, the detection of nearby individual WD mergers
would be of significant interest for understanding their physics and constraining the
merger rates. On the other hand, depending on the uncertainties, they could be a significant 
contribution to the diffuse neutrino background observed by IceCube 
in the range of $\sim 10\TeV$ to several PeV.

\section{Multi-Messenger Detection Programs: AMON}
\label{sec:multi}

Since charged cosmic rays arrive to us via a diffusive motion caused by scattering in the 
turbulent intergalactic and galactic magnetic field, their directionality is largely lost, 
and the corresponding time delays make it impossible to rely on time coincidences with
any time variability in the electromagnetic luminosity of the presumed sources. Thus,
the use of secondary radiations produced by the interactions of the cosmic rays, 
such as neutrinos or gamma-rays, provide an attractive, and perhaps the most immediate, 
way to identify and study the CR sources. 

\begin{figure}[ht]
\includegraphics[width=0.5\textwidth,height=2.0in,angle=0.0]{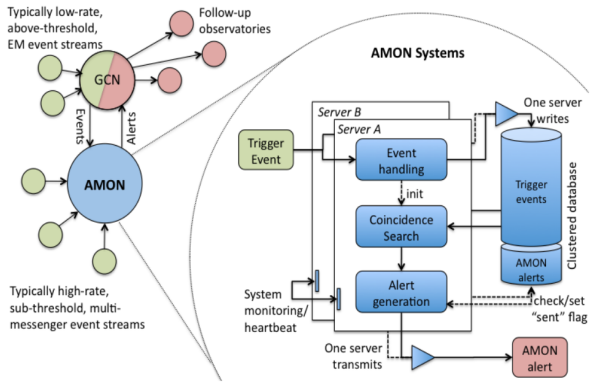}
\caption{
AMON manages data from multiple subthreshold multimessenger event streams (left), converting to 
a standard VOEvent format for storage in a secure clustered database (right). In addition, AMON
accepts above-threshold VOEvents from the Gamma-Ray Bursts Coordinate Network (GCN) and uses the 
GCN to distribute the AMON alerts.  From \cite{Smith+13amon}.
}
\label{fig:AMON-flowchart}
\end{figure}

There are numerous bi-lateral agreements between observatories which detect one or the other
of these possible secondary messengers at different energies, such as between IceCube and Swift, 
HAWC and Fermi, etc. There is however an ambitious observational program which serves as a centralized, 
{\bf multi-lateral} hub between a large number of disparate observing facilities,  called AMON 
\cite{Smith+13amon}, see Fig.  \ref{fig:AMON-flowchart}.

The AMON facility is a multi-institution program which has developed algorithms and codes
for interpreting triggers from simultaneous live alerts in two or more disparate messenger 
types supplied by different observatories and detectors.  The observatories that
participate in  AMON so far are ANTARES, FACT, Fermi LAT, Fermi GBM, HAWC, IceCube, LIGO, 
LMT, MASTER, Pierre Auger, PTF, Swift BAT, Swift XRT, Swift UVOT and  VERITAS. The algorithms 
are designed for  exploiting signals which are sub-threshold in individual observatories, 
but which can above-threshold signals when considered together with concurrent sub-threshold 
signals from other observatories. This generates an alert which is then re-distributed via 
internet to the participating observatories and observers. 

The system has transitioned in 2016 to real-time operations \cite{Cowen+16amon} and has now 
been online for several months. In the near future, it is hoped that such efforts may lead 
to important clues about the sources of UHECRs, very high energy neutrinos, very high energy 
gamma-rays and gravitational waves.

\section{Discussion}
\label{sec:disc}

The origin of the highest energy $\sim 10^{20}-10^{21}\eV$ cosmic rays remains the subject of 
intense debate, as does, for that matter, the origin of the CR spectrum down to at least 
$10^{15}\eV$. An interesting case can be made that the energy input rate per decade of energy
into the Universe $E^2 dN/dE \sim 10^{44}\erg\Mpc^{-3}\yr^{-1}$ is approximately constant from 
GeV to $\sim 10^{20}\eV$ \cite{Katz+13uhecr}. This would imply an approximate $N(E) \propto 
E^{-2}$ spectrum at all energies, suggesting a single type of source responsible for it,
although the nature of these sources is not known. As discussed in \S \ref{sec:grb}, GRBs
could be responsible at least for the $10^{19}-10^{21}\eV$ range, and indeed, if below $10^{19}
\eV$ the diffusion out of the host structures (galaxies, galaxy clusters, etc) results
in a steepening of the spectrum observed at Earth, GRBs might perhaps account for the whole range
down to $\sim$ TeV (the jet bulk Lorentz factor $\Gamma\sim 10^2=10^3$ resulting in an observed
lower limit around that energy). However other sources may also come into consideration, 
including AGNs, or, if the highest energy CRs are mainly heavy elements, hypernovae or galactic 
shocks. For the highest energies, however, the lack of steady and energetic enough sources
within the GZK radius is an argument in favor of transient sources.

For cosmic rays in the $10^{15}-10^{18}\eV$, a connection with  the observed diffuse
neutrino background observed by IceCube and the residual isotropic gamma-ray background 
observed by Fermi imposes constraints on possible models, as discussed in ~\S\S
\ref{sec:choked}, \ref{sec:tde} and \ref{sec:wd}. While  the astrophysical uncertainties
about the rates are substantial, electromagnetically dim (``hidden") sources such as choked 
GRBs (\S \ref{sec:grb}) could satisfy simultaneously the IceCube observations and the Fermi 
constraints; and for more optimistic assumptions, tidal disruption events (\S \ref{sec:tde})
or white dwarf mergers (\S \ref{sec:wd}) may also contribute.
Interestingly, the cosmic ray energy corresponding to the $\sim$ few PeV upper end of the 
observed neutrino energy is a few times $10^{17}\eV$, roughly corresponding to the second knee 
in the CR spectrum, and roughly in the energy range of a hypothesized (e.g. \cite{Gaisser+13crsp})
``third" spectral component" of CR sources. 

In summary, while cosmic rays and neutrinos are notoriously elusive preys, new observations 
throughout the next decade can be expected to lead to significant advances in our understanding 
of both UHECRs and their multi-messenger secondaries, as well as in the implications they bear
for the physical nature of their sources.

\noind
{\it Acknowledgments:} P.M. is grateful to the organizers of the 2nd ``Cosmic Ray Origin - 
Beyond the Standard Models" conference in San Vito di Cadore, Dolomites, Italy (2016); 
and to NASA NNX 13AH50G for partial support.




\bibliographystyle{elsarticle-num}


\end{document}